\documentclass[reprint, amsmath,amssymb, aps, prl, floatfix]{revtex4-2}

\usepackage{graphicx}
\usepackage{dcolumn}
\usepackage{bm}
\usepackage{hyperref}
\usepackage{epstopdf}
\usepackage{color}

\usepackage{graphicx,subfigure}
\usepackage{float}
\usepackage{color}
\usepackage{epstopdf}
\usepackage{amsmath}
\usepackage{amsthm}
\usepackage{amssymb}
\usepackage{amsfonts}
\usepackage{graphicx}
\usepackage{txfonts}
\usepackage[normalem]{ulem}
\usepackage{mathtools}
\usepackage{bm}
\usepackage{bbm}
\usepackage{amsmath}
\usepackage{graphicx}
\usepackage{ulem}
\usepackage{dcolumn}
\usepackage{epsfig}
\usepackage{bm}
\usepackage{array}
\usepackage{hyperref}
\hypersetup{
colorlinks=true,
citecolor=black,
linkcolor=red,
urlcolor=black, 
bookmarks=true,
pdfmenubar=true
}

\newcommand{\be}{\begin{equation}}
\newcommand{\ee}{\end{equation}}
\newcommand{\ud}{\mathrm{d}}
\newcommand{\beq}{\begin{eqnarray}}
\newcommand{\eeq}{\end{eqnarray}}
\newcommand{\Eins}{\ensuremath{\mathbbm 1}}

\def\ket#1{\left|#1\right\rangle}
\def\bra#1{\left\langle#1\right|}

\newcommand{\Tr}{\textrm{Tr}}
\newcommand{\mean}[1]{\overline{#1}}

\begin{document}

\title{Tomography of a number-resolving detector by reconstruction \\ of an atomic many-body quantum state}

\author{Mareike Hetzel$^1$} 
 \email{hetzel@iqo.uni-hannover.de} 
\author{Luca Pezz\`e$^2$}
\author{Cebrail P\"ur$^1$}
\author{Martin Quensen$^1$}
\author{Andreas H\"uper$^{1,5}$}
\author{Jiao Geng$^{3,4}$}
\author{Jens Kruse$^{1,5}$}
\author{Luis Santos$^6$}
\author{Wolfgang Ertmer$^{1,5}$}
\author{Augusto Smerzi$^2$}
\author{Carsten Klempt$^{1,5}$} 

\affiliation{$^1$Institut f\"ur Quantenoptik, Leibniz Universit\"at Hannover, Welfengarten 1, D-30167 Hannover, Germany \\ $^2$QSTAR and INO-CNR and LENS, Largo Enrico Fermi 2, 50125 Firenze, Italy \\ $^3$Key Laboratory of 3D Micro/Nano Fabrication and Characterization of Zhejiang Province, School of Engineering, Westlake University, 18 Shilongshan Road, Hangzhou 310024, Zhejiang Province, China \\ $^4$Institute of Advanced Technology, Westlake Institute for Advanced Study, 18 Shilongshan Road, Hangzhou 310024, Zhejiang Province, China \\ $^5$Deutsches Zentrum f\"ur Luft- und Raumfahrt e.V. (DLR), Institut f\"ur Satellitengeod\"asie und Inertialsensorik (DLR-SI), Callinstra{\ss}e 30b, D-30167 Hannover, Germany \\ $^6$Institut f\"ur Theoretische Physik, Leibniz Universit\"at Hannover, Appelstra{\ss}e 2, D-30167 Hannover, Germany}%



\date{\today}

\begin{abstract}
The high-fidelity analysis of many-body quantum states of indistinguishable atoms requires the accurate counting of atoms.
Here we report the tomographic reconstruction of an atom-number-resolving detector.
The tomography is performed with an ultracold rubidium ensemble that is prepared in a coherent spin state by driving a Rabi coupling between the two hyperfine clock levels.
The coupling is followed by counting the occupation number in one level. We characterize the fidelity of our detector and show that a negative-valued Wigner function is associated with it.
Our results offer an exciting perspective for the high-fidelity reconstruction of entangled states and can be applied for a future demonstration of Heisenberg-limited atom interferometry.
\end{abstract}

\maketitle

High-fidelity preparation, manipulation, and detection of quantum states of many indistinguishable atoms have been greatly improved during the last decades. 
These improvements facilitate exciting developments ranging from fundamental quantum atom optics experiments~\cite{CroninRMP2009, ByrnesBOOK} to entanglement-enhanced metrology applications~\cite{Pezze2018}.
In metrology, entangled states of neutral atoms serve as highly sensitive input states of atom interferometers, reducing the resolution limit from the Standard Quantum Limit to the ultimate Heisenberg limit~\cite{PezzePRL2009}.
Possible applications range from quantum interferometry~\cite{Appel2009,Schleier-Smith2010,Gross2010,Riedel2010,Lucke2011,Chen2011,Hamley2012,Berrada2013,StrobelSCIENCE2014,Hosten2016} and magnetometry~\cite{Wasilewski2010,Sewell2012,Muessel2014,Ockeloen2013} to atomic clocks~\cite{Louchet-Chauvet2010,Leroux2010b,KrusePRL2016,PedrozoNATURE2020} and inertial sensing~\cite{AndersPRL2021,Greve2021}.
To date, the atom counting noise represents one of the most crucial limitations in current experiments, affecting both fundamental studies and metrological applications.

\begin{figure}[ht!]
\centering
  \includegraphics[width=\columnwidth]{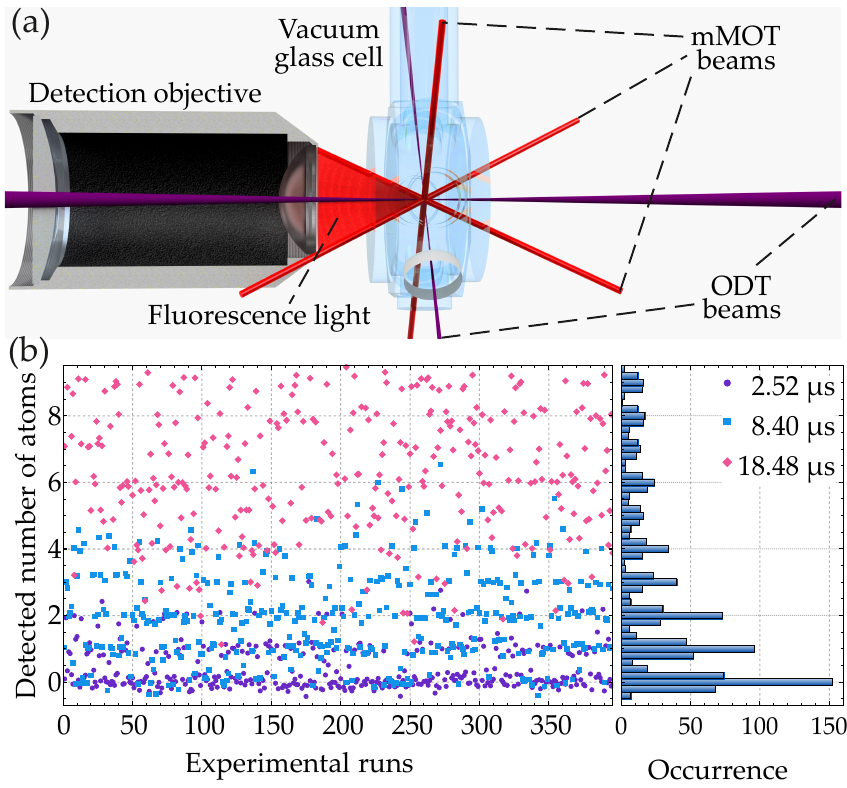}
    \caption{(a) Sketch of the experimental setup. 
    A coherent spin state is created in a crossed-beam optical dipole trap and analyzed in a number-resolving mMOT detection scheme. 
    (b) Time trace of the sequentially measured number of atoms in dependence of the MW pulse length. 
    The number of atoms in $\ket{1,0}$ after $2.52 ~\mu$s (purple circles), $8.4~\mu$s (blue rectangles) and $18.48 ~\mu$s (pink diamonds) is shown for up to 9 atoms for 1106 successive measurements. 
    The accumulation of data points at integer numbers is indicating our number-resolving counting. 
    The histogram of the fluorescence signal on the right further illustrates this effect.
    Negative values are caused by background substraction.
  \label{fig1}
  }
\end{figure}

Recent experiments creating entangled atomic quantum states in ensembles of indistinguishable atoms have reported counting noise that ranges from 3~atoms at a total number of a 600 atoms~\cite{Fadel2018}, to 1.6~atoms at $3000$ atoms~\cite{Qu2020}, to 10~atoms at $10^4$~\cite{Lucke2014,Luo2017}, to better than 17~atoms at $10^5$~\cite{Hosten2016}, and to 50~atoms at $5 \times 10^5$~\cite{Hamley2012}. 
An improvement of the counting noise below the value of the single atom, where the quantization of the atomic signal becomes apparent, promises a quantitative and qualitative improvement.
For example, such a counting resolution would allow for the direct detection of Bell correlations between two separated atomic ensembles~\cite{Laloee2009} and the observation of parity signals in Hong-Ou-Mandel-like interference experiments with many-particle states~\cite{CamposPRA1989, OuPRL1999}.
In metrology, a number-resolving counting can be applied to demonstrate a Heisenberg-limited resolution in atom interferometry~\cite{Holland1993,Bouyer1997}.
A number-resolving counting has been obtained in a cavity-based detection~\cite{Haas2014}, but is so far restricted to a discrimination between 0 and 1, and the scaling to larger numbers is an open challenge.
Single-atom resolved detection has also been obtained in free-falling clouds with a sheet of resonant light~\cite{Buecker2009}, and was applied to extract local correlations.
A number-resolved counting for up to 1000 atoms has been demonstrated in a millimeter-sized magneto-optical trap (mMOT)~\cite{Hume2013,Stroescu2015}, but was so far not applied to the detection of many-body quantum states.

\begin{figure*}[ht!]
\centering
  \includegraphics[width=\textwidth]{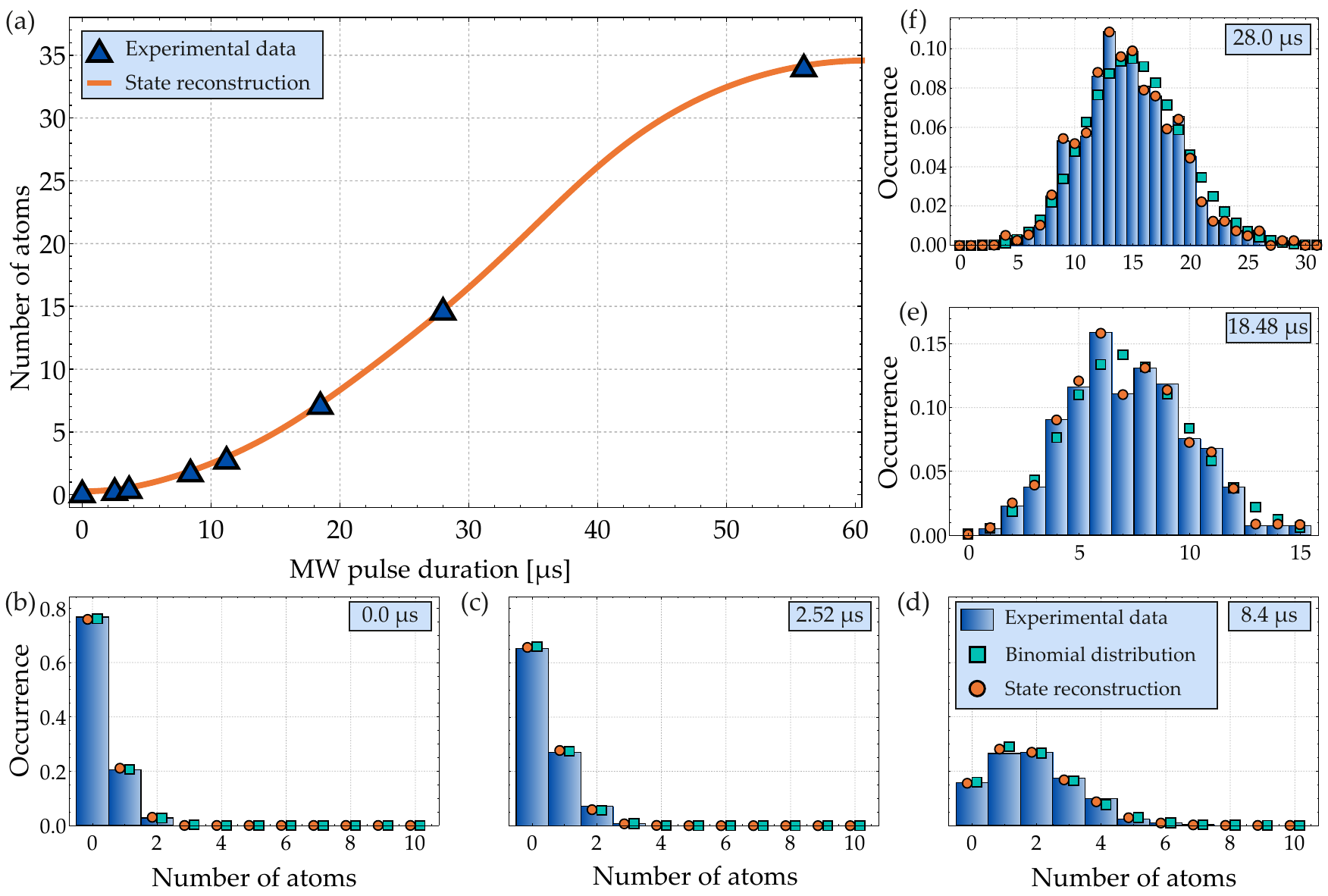}
    \caption{ (a) Mean number of atoms detected in $\ket{1,0}$ in dependence of the MW pulse duration. 
    Each data point (blue triangle) corresponds to an individual atom number distribution. 
    Those are exemplarily shown in the histograms in (b)-(f) for MW pulses ranging from $t=0\,\mu$s to $28\,\mu$s. 
    The ideal binomial distributions accounting for the detection offset and the atom number fluctuations are illustrated in the cyan rectangles.
    The solid orange line and the orange circles show the results obtained from the QDT algorithm (see text).
  \label{fig2}
  }
\end{figure*}

The fine calibration of quantum measurement devices generally requires quantum detection tomography (QDT) techniques~\cite{LuisPRL1999, FiurasekPRA2001}. 
The goal of QDT is to provide a set of positive-operator-valued measures (POVM) that fully characterize the detector, beyond the assumption of projective measurements.
So far, QDT has been mainly investigated for optical photocounting and homodyne detection~\cite{LundeenNATPHYS2009, ZhangNATPHOT2012, GrandiNJP2017, ZhangPRL2020, BridaNJP2012} and very recently also applied to characterize qubit readout for pairs of trapped ions~\cite{KeithPRA2018} and  quantum computing machines~\cite{ChenPRA2019}.
Although QDT is necessary for quantum state preparation, control, reconstruction, and error mitigation, its potential has not yet been leveraged for the characterization of neutral-atom quantum systems.
In this case, the technique is specifically promising, because of the small inherent detection loss compared to optical systems and the large achievable atom numbers compared to ion systems.

In this Letter, we apply a MOT-based number counting~\cite{Hueper2021} to analyze the dynamics of a many-body spin state.
We generate an atomic Bose-Einstein condensate (BEC) in one atomic clock level, apply a microwave (MW) coupling pulse of variable duration on the atomic clock transition, and count the number of atoms in the other, initially empty level.
We are able to follow the time evolution of the coherent spin state with a clear resolution of the number quantization.
By applying a stochastic matrix approach to the recorded histograms, we  obtain a set of nonclassical POVM operators that fully characterize the detection process.
The expected Poissonian distributions can be reproduced with a statistics-limited fidelity of up to 99$\%$.
We predict that the single-level detection is capable of operating an interferometric measurement with up to 7.8~dB squeezing-enhanced phase sensitivity gain, if the same total number can be provided with smaller fluctuations.
This sensitivity gain can be further enhanced by increasing the mean atom number.
Our detection capability promises a novel quality for the analysis of entangled quantum states and demonstration of Heisenberg-limited interferometry.

In our experiments, we generate a BEC of $10^5$ $^{87}$Rb atoms in a crossed-beam optical dipole trap with a preparation time of $3.3$~s. The details of the BEC production can be found in Ref.~\cite{accompanyingPRA}.
We prepare the BEC in the hyperfine level $\ket{F,m_F}=\ket{2,2}$ and reduce the number of atoms to enter the regime of our number-resolved counting.
The reduction is realized by transferring $34$~atoms to the level $\ket{1,1}$, on average, and a subsequent optical removal of the residual atoms in the $F=2$ manifold. 
A further MW pulse transfers the remaining atoms to the level $\ket{2,0}$.
A final resonant light push on the $F=1$ manifold terminates our state preparation with $34$ and $0$~atoms in the clock states $\ket{2,0}$ and $\ket{1,0}$, respectively.
The total number of particles fluctuates by $6.4$ atoms, which is dominated by projection noise ($5.8$ atoms).

We apply a resonant MW pulse on the clock transition with a variable duration ranging from $t=2.5\,\mu$s to $56\,\mu$s.
The many-body state in the pseudo-spin-1/2 system can thus be represented by a coherent spin state (CSS), with maximal total spin, but variable rotation angle $\theta$.
The analysis of the CSS is based on counting the number of atoms in the level $\ket{1,0}$.
To this end, the atoms in the level $\ket{2,0}$ are removed and the remaining atoms are counted by fluorescence detection in the mMOT.
The quality of the state analysis thus depends on the efficiency of the removal procedure.
Therefore, the detection process starts with a strong reduction of the atomic density by switching off one of the two dipole trap laser beams.
A $\sigma ^+$-polarized light push at a magnetic field of $6.7$~G quickly pumps all $F=2$ atoms into a closed cycling transition reducing the probability to fall into a non-resonant state.
The resonant atoms are accelerated and leave the trap, while the probability of unwanted collisions is reduced by the low density.
The removal of atoms in the level $\ket{2,0}$ has a finite extinction ratio of $42.4$~dB, resulting in an unwanted, Poisson-distributed remainder of $0.27$ atoms maximally.
These atoms are produced by two processes:
(i) They escape the removal process to the level $F=1$ because of imperfect optical pumping.
(ii) They are captured from the background gas, which is temporally increased after the operation of the two-dimensional magneto-optical trap. 
We detect the remaining atoms in the mMOT setup, consisting of a magneto-optical trap with millimeter-sized illumination beams~\cite{accompanyingPRA, Hueper2021}.
The optical dipole trap is switched off to start an equilibration phase in the mMOT of $50$~ms, during which the magnetic fields settle and the atoms are compressed and cooled.
Subsequently, the main atom counting signal is obtained by collecting  fluorescence light for $65$~ms with a charge-coupled-device (CCD) camera.
Finally, a second image without atoms is recorded for background subtraction.
The spin preparation and detection processes require a total of $1.8$~s.
After nine measurement runs, the system is halted for 60~s to avoid a slow increase of the mMOT capture rate from the background gas. 

Fig.~\ref{fig1} (a) shows a sketch of the experimental setup including the mMOT and ODT beams and the high-numerical-aperture detection objective.
Fig.~\ref{fig1} (b) shows a time trace of 100 consecutive number measurements in $\ket{1,0}$ shown for three different MW pulse lengths. 
The measured number of atoms accumulate at integer numbers, enabling a number assignment fidelity ranging from 99.7\% at 1 atoms to 99.0\% at 15 atoms~\cite{accompanyingPRA}.

Fig.~\ref{fig2} (a) shows the mean number of the transferred atoms as a function of the microwave pulse duration.
The mean atom number follows a sinusoidal Rabi oscillation with a Rabi frequency $\Omega= 2 \pi \times 8.2$~kHz (see below).
Figs.~\ref{fig2} (b)-(f) present the exemplary histograms, which can be associated to rotation angles $\theta=\Omega t$.
Without rotation (b), the distribution shows the detection of recaptured atoms, which can be treated as statistical dark counts in the detection system.
For finite rotations (c)-(f), the distributions shift to higher atom number and increased width.

\begin{figure*}[ht!]
\begin{center}
\includegraphics[width=\textwidth]{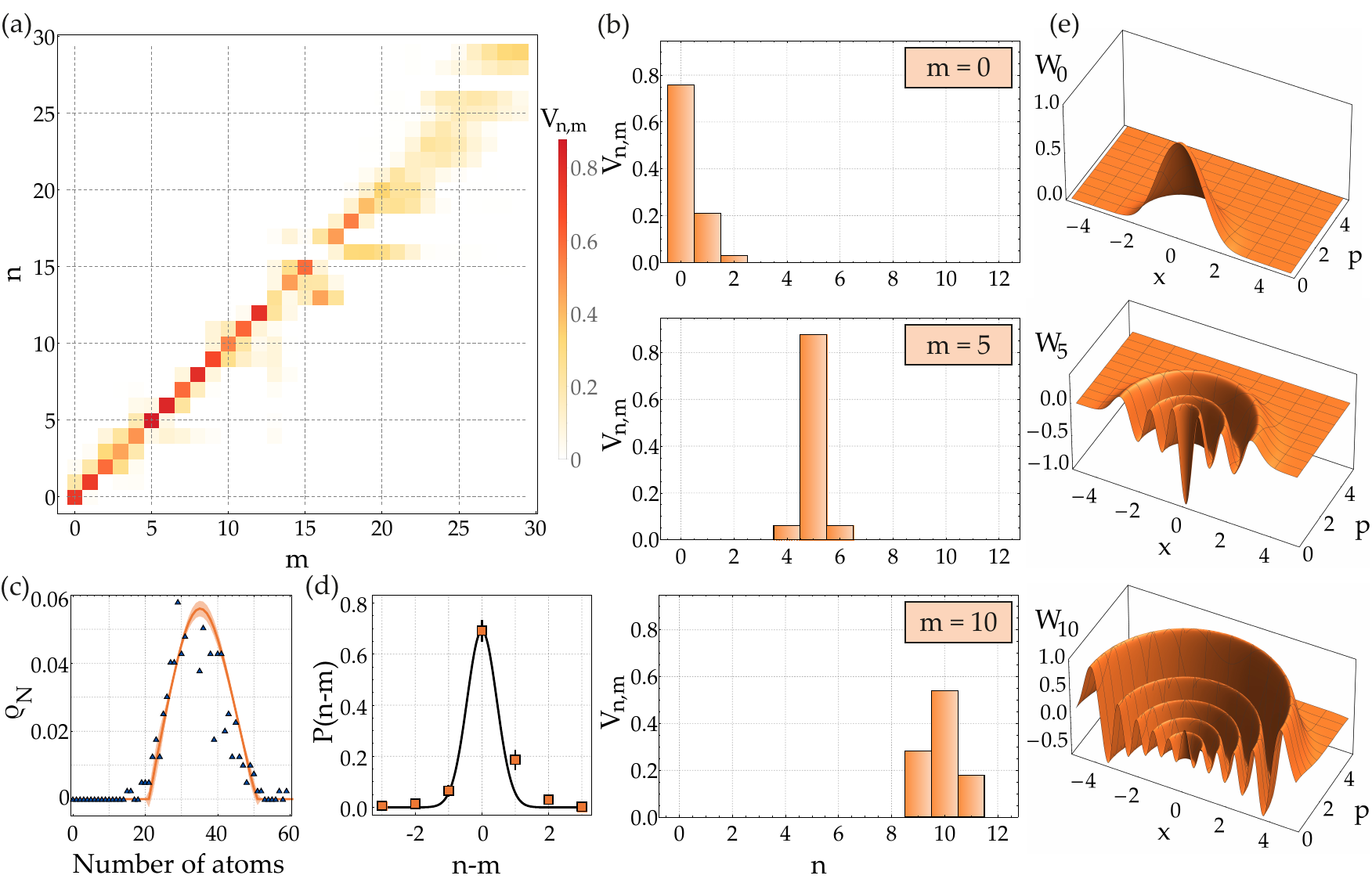}
\end{center}
\caption{Panel (a): reconstructed stochastic map $V_{n,m}$ (linear color scale) as a function of $n$ and $m$. 
The highest weight is concentrated along the diagonal $n=m$.
Panel (b): $V_{n,m}$ as a function of $n$ and for $m=0$, $m=5$ and $m=10$.
Panel (c): coefficients $\rho_N$ of the reconstructed state (orange line with uncertainty shade, see text). 
The blue triangles show the experimentally obtained state in $\ket{1,0}$ after a $56\,\mu$s MW pulse.
Panel (d) shows $P(n-m)$ as a function of $n-m$ (squares). 
The black line is a Gaussian fit to the data for $n\leq m$
Panel (e): the Wigner function of the POVM operator $\hat{\Pi}_{0}$, $\hat{\Pi}_{5}$ and $\hat{\Pi}_{10}$, in the $(x,p)$ phase space.
Negative Wigner values are observed for $\hat{\Pi}_{n\geq 1}$.
}  
\label{Fig3}
\end{figure*}

Under the assumptions that the microwave generates a homogeneous coupling to the cloud and that the level $\ket{1,0}$ is initially completely empty -- that are both very well fulfilled in our case -- we can employ the recorded data for QDT.
We associate the two clock levels $\ket{1,0}$ and $\ket{2,0}$ with the letters $a$ and $b$, respectively, to simplify the notation.
We model the detection by expressing the probability of a measurement result $n$ (the number of atoms in clock level a) as 
\be \label{ProbV}
P_V(n \vert t) = \sum_{m=0}^{+ \infty} V_{n,m} P_{\rm id}(m\vert t),
\ee
in terms of a stochastic matrix (Markov mapping) $V$ with non-negative elements $V_{n,m} \geq 0$, which satisfies the normalization property $\sum_{n} V_{n,m} = 1$ for all $m$.
Physically, the quantity $V_{n,m}$ can be interpreted as the probability to measure $n$ atoms if $m$ atoms reach the detector.
We use the ideal probability $P_{\rm id}(m\vert t) = \Tr[\vert m \rangle \langle m \vert \hat{\rho}(t)]$, where $\hat{\rho} = \sum_{N=0}^{+\infty} \rho_N \ket{N}_b \ket{0}_a \bra{0}_a \bra{N}_b$ is the generic atomic state before starting the dynamics, $\hat{U}(t) = \exp \big[-i \Omega_R t (\hat{a}^\dag \hat{b} + \hat{a} \hat{b}^\dag)/2 \big]$ describes the Rabi coupling and $\hat{\rho}(t) = \hat{U}(t) \hat{\rho} \hat{U}(t)^\dag$.
The assumption that the initial state is diagonal is well justified experimentally.
The matrix $V$ provides a full characterization of the detection process, including finite resolutions and biases.
It should be noticed that Eq.~(\ref{ProbV}) can be rewritten as $P_V(n \vert t) = \Tr\big[\hat{\rho}(t) \hat{\Pi}_n\big]$, in terms of a POVM set $\{ \hat{\Pi}_n \}$, where
\be \label{Eq.POVM}
\hat{\Pi}_n = \sum_{m=0}^{+ \infty} V_{n,m} \vert m \rangle \langle m \vert.
\ee
$V$ being positive semi-definite guarantees that $\hat{\Pi}_n\geq 0$, while the condition $\sum_{n} V_{n,m} = 1$ for all $m$ guarantees the completeness relation $\sum_n \hat{\Pi}_n = \Eins$.

Our QDT protocol consists of finding the coefficients $\rho_N$, $V_{n,m}$ and $\Omega_R$ that minimize a cost function $\mathcal{C} = \sum_j d_H^2(t_j)$.
This is given by the sum over all times $t_j$ of the squared statistical distance $d_H^2(t_j)$~\cite{BengtssonBOOK} between the probability distribution $P_V(n \vert t_j)$ and 
the experimental histogram $P_{\rm exp}(n \vert t_j)$ [e.g. Fig.~\ref{fig2}(b)-(f)],
\be \label{Eq.fid}
d_H^2(t_j) = \sum_{n} \Big( \sqrt{P_V(n \vert t_j)} - \sqrt{P_{\rm exp}(n \vert t_j)} \Big)^2.
\ee
The constrained minimization of Eq.~(\ref{Eq.fid}) is performed with a  gradient descent algorithm~\cite{supplements}.
We emphasize that a reliable characterization of the POVM set requires a substantial overlap between probability distributions at different times in order to avoid overfitting~\cite{supplements}. 
For our experimental parameters, see Fig.~\ref{fig2}, this is guaranteed for $0 \leq n \lesssim 20$.

In Fig.~\ref{fig2} we compare the experimental histograms (bars) with the probabilities derived from the QDT (orange circles), namely Eq. (\ref{ProbV}) with $V$, $\rho_N$ and $\Omega_R$ calculated using the minimization algorithm.
The agreement is excellent, as the obtained probability distribution $P_V(n \vert t_j)$ achieves a very high fidelity with $P_{\rm exp}(n \vert t_j)$, for all $t_j$ (notice that the iterative optimization algorithm is stopped when $\mathcal{C} = 0.01$, which is a value close to saturation~\cite{supplements}).
The histograms are consistent with that calculated with a binomial distribution (green squares).
The latter assume a Gaussian distribution of the total number of atoms with measured mean and standard deviation, ideal Rabi transfer and the convolution with a binomial distribution with a mean number of 0.27~atoms to account for the unwanted detection of background atoms.
The mean number of atoms as a function of time, $\sum_n P_V(n,t) n$ for the reconstructed $V$, $\rho_N$ and $\Omega_R$ interpolates well the detection events as shown in Fig.~\ref{fig2}(a).
It should be noticed that the Rabi frequency extracted from the tomographic reconstruction $\Omega_R = 8.2 \pm 0.2$~kHz agrees precisely with the result of a sinusoidal fit to the data.

In Fig.~\ref{Fig3} we show the results of our joint detection and state reconstruction. 
Figure~\ref{Fig3}(a) shows the elements $V_{n,m}$.
For most values of $m$, the weights $V_{n,m}$ concentrate around the diagonal $n=m$, where they reach their maximal value.
In other words, if $m$ atoms reach the detector, the most probable event is to detect $n=m$.
The probability of such detection events is quantified below.
For instance, in panel~(b) we show $V_{n,m}$ as a function of $n$ and for the specific values $m=0$, $5$ and $10$: the histograms are cuts of the plot of panel (a).
For $m \gtrsim 20$, the reconstructed $V_{n,m}$ spreads away from the diagonal.
Here, the QDT becomes uncertain because the recorded probability distributions do not overlap sufficiently and the optimization method is affected by overfitting of the data \footnote{See for instance M. A. Nielsen, {\it Neural Networks and Deep Learning} (Determination Press, 2015), available online at http://neuralnetworksanddeeplearning.com}.
To recognize the overfitting effect, we have performed a 'learning test', see Ref.~\cite{supplements}, in which the experimental histogram at time $t_j$ is compared with the reconstructed $P_V(n\vert t_j)$, where the coefficients $\Omega_R$, $\hat{\rho}$, and $V$ are calculated from the minimization algorithm using all the experimental data except those at time $t_j$.
In this case, we observe a fidelity between $P_V(n\vert t_j)$ and $P_{\rm exp}(n\vert t_j)$ above $99\%$ for times $t_j$ up to 18.48 $\mu$s~\cite{supplements}.
In the future, a QDT at larger atom numbers can be obtained by taking more histograms with larger statistics.
In panel (c) we show the reconstructed elements $\rho_N$ as a function of the number of particles.
As we see, the reconstructed diagonal state has approximately a Gaussian shape with mean $\mean{N} = 35.4$ and root mean square error $\Delta N = 6.4 \approx \mean{N}^{1/2}$.

For $m \lesssim 20$, as $V_{n,m}$ is strongly peaked around $n=m$.
It is thus convenient to calculate $P(n-m) = \sum_{m} V_{n-m,m} P_{m}$, giving the reconstructed probability that $n-m$ particles are detected if $m$ particles hit the detector.
Here, $P_{m} = \sum_{j} P_{\rm id}(m \vert t_j)$ is the overall probability that $m$ particles hit the detector, for the considered measurement times $t_j$ and takes into account the most likely detection events and it is almost negligible for $m \gtrsim 20$.
For a noiseless detector $P(n-m)$ is a delta peak at $n=m$, regardless the $P_m$ distribution. 
In the case of our noisy detector, $P(n-m)$ is still strongly peaked at $n=m$, with an overall probability of about $70 \%$, see  Fig.~\ref{Fig3} (d).
The slight asymmetry of the distribution $V_{n-m}$ reflects the unwanted recapture of atoms described above, which biases $V_{n-m}$ to positive values of $n-m$.
By calculating the variance of the $P(n-m)$ distribution for $n \leq m$ (thus not affected by the atom recapture) we can extract a detection sensitivity $\sigma = 0.4 \pm 0.02$.
This counting uncertainty is larger than the uncertainty obtained from Fig.~\ref{fig1}, because the finite number of measurements additionally deteriorates the QDT.

As shown in Eq.~(\ref{Eq.POVM}), accessing the matrix $V$ allows us to characterize the POVM elements $\hat{\Pi}_n$. 
For instance, in Fig.~\ref{Fig3}(d), we plot the Wigner distribution of the reconstructed POVM operators $\hat{\Pi}_n$, $W_{n}(x,p) = \sum_{m} V_{n,m} W(x,p;m)$, where $W(x,p;m)$ is the Wigner function of the Fock state  $\ket{m}$~\footnote{
Explicitly, $W_{n}(x,p) =
\sum_{m} V_{n,m} \frac{(-1)^m}{\pi} e^{-(x^2 + p^2)} L_m\big[2(p^2 + x^2)\big]$,
where $L_m(x)$ denotes the $m$-th Laguerre polynomial.}.  
For $n=0$ the Wigner function is positive, as expected, corresponding to the detection of vacuum. 
On the contrary, for $n \geq 1$, the Wigner functions $W_{n}(q,p)$ have negative values, indicating the absence of a classical analogue of these operators. 
To emphasise the fundamental quantum nature of our detection we notice that a POVM with negative Wigner function is necessary to prove Bell's non-locality with Gaussian states (which have positive Wigner distributions)~\cite{LundeenNATPHYS2009,BanaszekPRA1998}.

Measuring the atom number in a single level, as done in our experiment, still allows to surpass the standard quantum limit of phase sensitivity, for instance provided that (i) each fixed-$N$ state is spin squeezed and (ii) the distribution of the total atom number $\rho_N$ has sufficiently low fluctuations. 
Using the results of our QDT and the atom number distribution, we can predict an optimal gain of about 1.3 over the standard quantum limit, when squeezing the relative atom number distribution~\cite{supplements}. 
A higher gain, up to 7.8~dB is possible when also reducing $\Delta N$, for our detection and $\bar{N} \approx 36$.
The gain further increases when increasing $\bar{N}$.

In summary, we have employed a number-resolving detector to analyze the dynamics of a coherent spin state derived from an atomic BEC.
We have characterized the detection process by the simultaneous reconstruction of the diagonal quantum state and the detector's POVM operators.
The latter are characterized by negative Wigner functions, thus unveiling the inherent quantum nature of the detector.
In the future, the presented detector and the developed QDT techniques will be directly extended to entangled many-body states, promising the detection of entanglement with unprecedented fidelity in the regime of up to 100 atoms.
 
\begin{acknowledgements}
  {\it Acknowledgements.} 
  This work is supported by the QuantERA grants SQUEIS and MENTA.
  We acknowledge financial support from the Deutsche Forschungsgemeinschaft (DFG, German Research Foundation)-Project-ID 274200144-SFB 1227 DQ-mat within the project B01 and Germany’s Excellence Strategy—EXC-2123 QuantumFrontiers—Project-ID 390837967. 
  M.Q. acknowledges support from the Hannover School for Nanotechnology (HSN).
\end{acknowledgements}

\bibliography{main,Luca}

\pagebreak
\clearpage
\widetext
\begin{center}
\textbf{\large Tomography of a number-resolving detector by reconstruction of an atomic many-body quantum state (Supplemental Material)}
\end{center}
\setcounter{equation}{0}
\setcounter{figure}{0}
\setcounter{table}{0}
\makeatletter
\renewcommand{\theequation}{S\arabic{equation}}
\renewcommand{\thefigure}{S\arabic{figure}}

In the Supplemental Material we provide more details about the optimization algorithm for the quantum  detection  tomography as well as a detailed analysis about the possibility to reach sub-shot noise sensitivities with the noisy detector and single-level counting.  


\section{Quantum detection tomography}

\subsection{Optimization algorithm}

The minimization of the cost function $\mathcal{C}= \sum_{j} d_H^2(t_j)$, with $d_H^2(t_j)$ given by Eq. (\ref{Eq.fid}) of the main text, is done using a gradient descent algorithm.  
The gradient descend technique is a widely used method to find the minimum of a cost function with respect to a large number of free parameters. 
It is typically used in neural networks and variational quantum algorithm.
Starting from an initial guess, the parameters are updated iteratively by following the direction of the negative gradient of $\mathcal{C}$.
Notice that, in place of $d_H^2(t_j)$ in the cost function, we could have used the the Kullback–Leibler divergence between $P_V(n \vert t_j)$ and  $P_{\rm exp}(n \vert t_j)$, obtaining similar results. 

In our case, the minimization is divided in different steps. 
The numerical optimization provides the stochastic matrix $V,$ the coefficients $\rho_N$ and the Rabi frequency $\Omega_R$. 
First, we run a number of iterations to minimize $V$, followed by a minimization of $\rho_N$, and finally, of $\Omega_R$. 
The procedure is then repeated iteratively until $\mathcal{C}= 0.01$, which is set as a convenient cutoff.
This cutoff is very close to saturation, meaning that increasing the number of optimization steps of the algorithm, $\mathcal{C}$ can reach values slightly smaller than 0.01 at the price of an rapidly increasing computation time.  

The minimization algorithm does not require any input parameter.
It relies on assuming the initial state in the diagonal form (see main text) and that the transformation follows a unitary evolution. 
Both these assumptions are well justified experimentally.
The number of particles domain is cut to the maximum detected value, $n_{\rm max}$ (equal to 59 in our case), such that $V$ is a square matrix of dimension $(n_{\rm max}+1)^2$.
We impose the following constraints at each step of the algorithm: i) $V \geq 0$; ii) $\sum_{n} V_{n,m} = 1$ for all $m$; iii) $\rho_N \geq 0$ for all $N$; and iv) $\sum_N \rho_N = 1$.
As initial conditions we take a {\it random} and normalized $(n_{\rm max}+1) \times (n_{\rm max}+1)$ matrix $V$, a Gaussian distribution for $\rho_N$ and an initial guess for $\Omega_R$.
The center of the Gaussian number distribution and the Rabi frequency are extracted from a fit of the mean. number of atoms $\mean{N}(t)$ as a function of time $t$, according to the function $\mean{N}(t) = \mean{N} \sin^2 (\pi \Omega_R t) + C_{\rm off}$, with $\mean{N}$, $\Omega_R$ and the offset $C_{\rm off}$ as fitting parameters.
The width of the Gaussian, $\Delta N$ is extracted from a fit of $\mean{N^2}(t)$ as a function of time according to $\mean{N^2}(t) = \sum_{k=0}^4 b_k \sin^k (\pi \Omega_R t)$, where $b_k$ are fitting parameters.
We identify $b_4 = \mean{N^2}$. 
Notice that for these initial conditions the initial cumulative fidelity is very small, $\mathcal{F} \approx 0.5$.
We have verified that the output of the gradient descent algorithm does not depend on the choice of the initial random matrix $V$.
 
Error bars on quantities of interest are obtained from a resampling of the experimental probability distribution. 
Specifically, given the reconstructed $V$, $\rho_N$, and $\Omega_R$, we generate about 20 new sets of $n_{\rm max}$ resampled data for each time $t_j$ ($j=1, ...,8$),
mimicking re-acquisitions of experimental data sets.
For each data set we run the gradient descent optimization.
Overall, this procedure provides a set of 20 matrices $V$, $\rho_N$, and $\Omega_R$, from which we extract error bars.

\subsection{Learning test}

\begin{figure}[h!]
\begin{center}
\includegraphics[width=1\columnwidth]{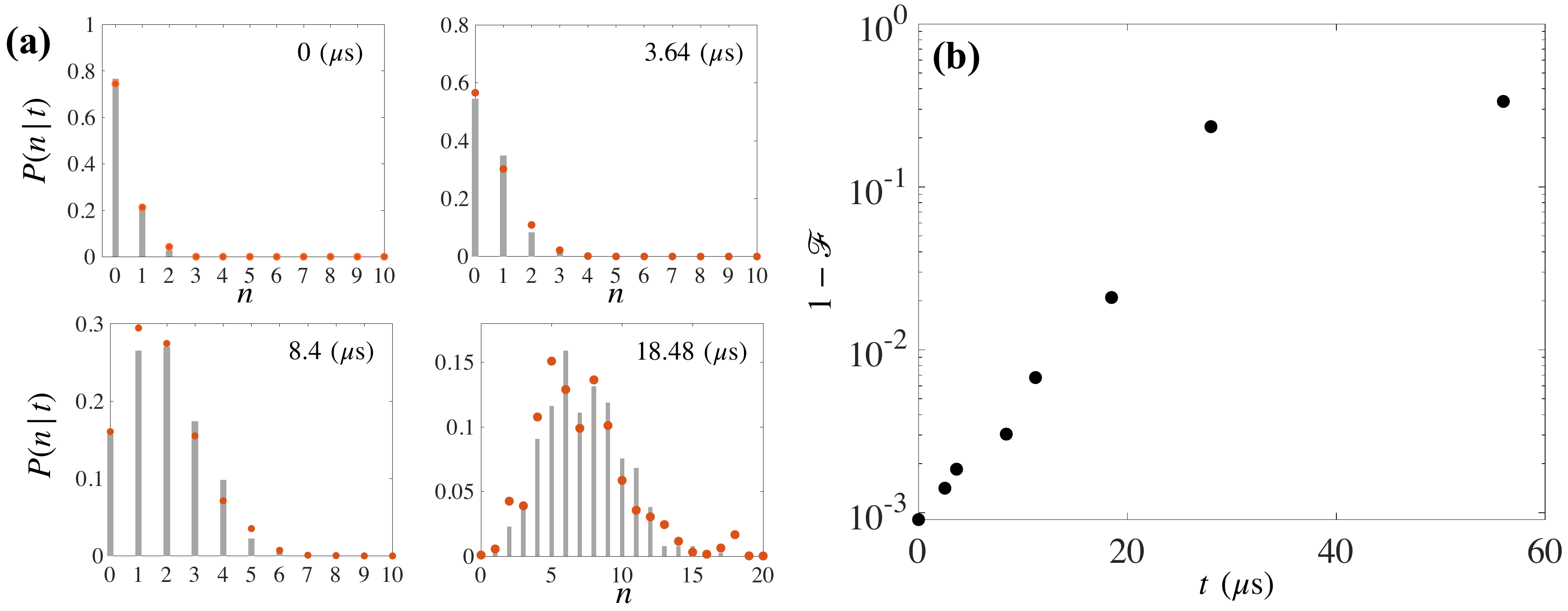}
\end{center}
\caption{Learning test. We apply the tomography method using all data expect at time $t_j$ in order to guess the experimental distribution at time $t_j$.
(a) The learned distribution $P_{V_j}(n\vert t_j)$ (red circles) is compared to the experimental distribution (grey bars), at different times. 
In panel (b) we show $1-\mathcal{F}(t_j)$, where $\mathcal{F}(t_j)$ is the fidelity between the experimental distribution $P_{\rm exp}(n, t_j)$ and $P_{V_j}(n\vert t_j)$, Eq.~(\ref{Fidj}).
}  
\label{FigSupp1}
\end{figure}

Here we provide a "learning test'' of our method.
For this, we reconstruct $\Omega_R$, $\hat{\rho}$, and $V$ using the experimental data collected at all time {\it except} at time $t_j$.  
Numerical methods and initial conditions are the same as in the previous analysis. 
The reconstructed quantities (labelled with index $j$,
$\Omega_R^{(j)}$, $\hat{\rho}^{(j)}$, and $V^{(j)}$) are used to calculate the expected output probability $P_{V_j}(n\vert t_j)$ at the time $t_j$ not seen by the minimization algorithm. 
In Fig.~\ref{FigSupp1}(a) we compare the probability $P_{V_j}(n\vert t_j)$ (red circles) with the experimental histograms $P_{\rm exp}(n\vert t_j)$ (grey bars), at different times.
While at small times (e.g. at $t=0$ and $t=3.64$ $\mu$s), the method is able to learn the unseen distribution, this capability is lost for large times. 
This is due to the fact that, at large times, distributions do not overlap substantially: the algorithm simply ``overfits'' the experimental distributions without been able to generalize to unseen distributions.   
To be more quantitative, we calculate the fidelity
\be \label{Fidj}
\mathcal{F}(t_j) = \sum_{n}  \sqrt{P_j(n \vert t_j)} \sqrt{ P_{\rm exp}(n \vert t_j) },
\ee
between the experimental distribution and the reconstructed one. 
Results are shown in Fig. \ref{FigSupp1}(b).

\subsection{Fisher information}
\begin{figure}[t!]
\begin{center}
\includegraphics[width=0.45\columnwidth]{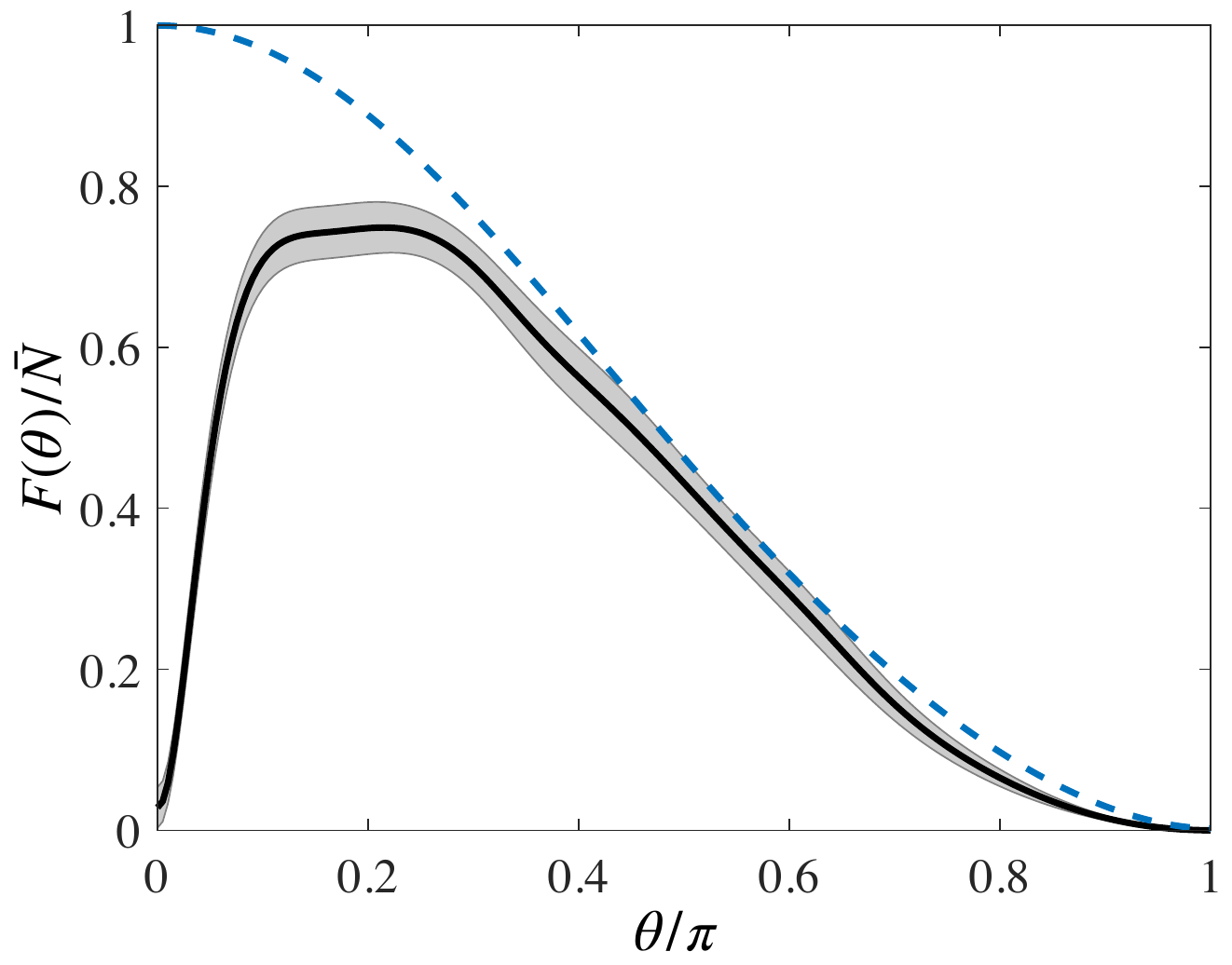}
\end{center}
\caption{Fisher information Eq.~(\ref{Fisher}) obtained from the reconstructed probability distribution (black line) as a function of $\theta$.
The grey region is the uncertainty extracted from bootstrapping. 
The blue dashed line is the Fisher information calculated using ideal probabilities, $F_{\rm id}(\theta)/\mean{N} = \cos^2(\theta/2)$.
}  
\label{FigSupp2}
\end{figure}

As a further consistency check of our results, we calculate -- using the reconstructed distribution $P_V(n\vert t)$ -- the Fisher information with respect to the parameter $\theta = \Omega_R t$:
\be \label{Fisher}
F_V(\theta) = \sum_{n} \frac{1}{P_V(n\vert \theta)} \bigg( \frac{\ud P_V(n\vert \theta)}{\ud \theta} \bigg)^2.
\ee
The quantity $F_{V}(\theta)/\mean{N}$ is shown by the solid black line in Fig.~\ref{FigSupp2}.
The grey region are fluctuations obtained via bootstrapping. 
The Fisher information is easily calculated using ideal probabilities, giving $F_{\rm id}(\theta)/\mean{N} = \cos^2(\theta/2)$ and shown as dashed blue line in the figure. 
Equation (\ref{Fisher}) follows the ideal behaviour for relatively large values of $\theta$, where the influence of (small) detection noise is less relevant.

\section{Possible observation of sub shot noise sensitivities}

The aim of this subsection is to evaluate the possibility to reach a sub-shot noise phase sensitivity by using i) a superposition of spin squeezed states, ii) measuring the number of particles in one atomic level, and iii) using the noisy detectors characterized in this work.
We model the probe state as a superposition, 
\begin{equation} \label{rhostate}
    \hat{\rho} = \sum_N \rho_N \,\vert \psi(s,N) \rangle \langle \psi(s,N) \vert, 
\end{equation}
of Gaussian states with different number of particles, 
\begin{equation}
    \vert \psi(s,N) \rangle = \mathcal{N}(N) \sum_{\mu=-N/2}^{N/2}  e^{-\mu^2/(4Ns)} \, \vert \mu \rangle_x    
\end{equation}
where $\mathcal{N}(N)$ provides the normalization and $\vert \mu \rangle_x$ are eigenstates of the operator $\hat{J}_x$ with eigenvalue $\mu$.
Here $\hat{J}_x=(\hat{a}^{\dag}\hat{b} + \hat{b}^{\dag}\hat{a})/2$ and $\hat{a}$ ($\hat{b}$) are annihilation operators for the two modes $a$ and $b$ corresponding to two atomic hyperfine levels.
The quantity $s$ is a parameter directly related to the variance of $\hat{J}_x$: 
$4 (\Delta \hat{J}_x)^2 = s \sum_N \rho_N N$.
The case $s=1$ approximates well, up to corrections of the order $1/N$, a superposition of coherent spin states. 
When $s<1$, each state $\vert \psi(s,N) \rangle$ is spin squeezed.
Following the experimental results, here we model $\rho_N$ as a Gaussian distribution with mean $\bar{N}$ and root mean-square fluctuations $\Delta N$ (we recall that $\bar{N}\approx (\Delta N)^2 \approx 36$ in our experiment). 

\begin{figure*}[h!]
\begin{center}
\includegraphics[width=\textwidth]{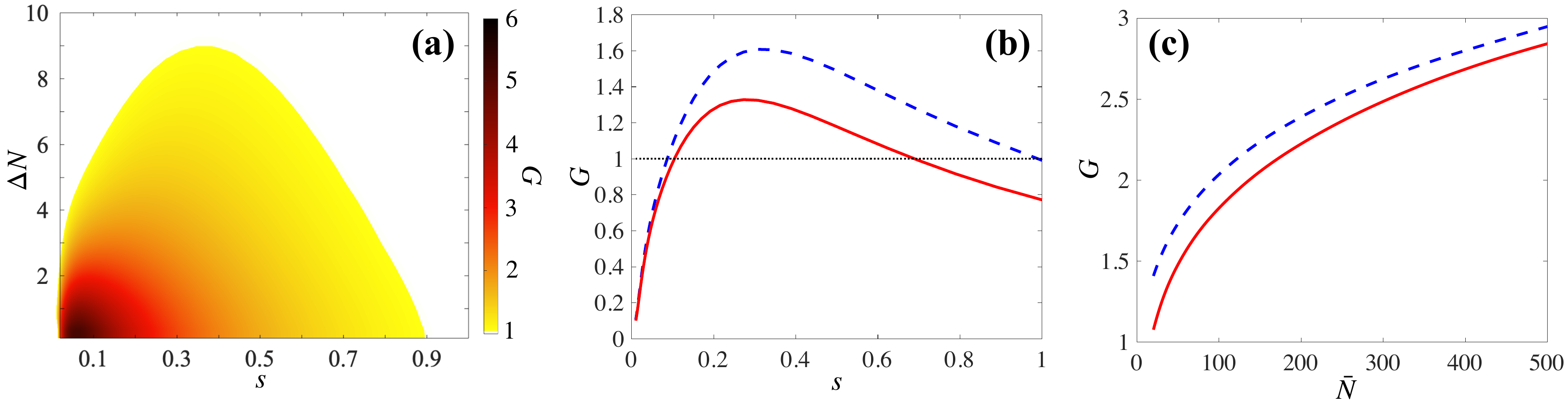}
\end{center}
\caption{Panel (a): phase sensitivity gain as a function of the squeezing parameter $s$ and the number of particles fluctuations, $\Delta N$.
Here $\mean{N}=36$.
Panel (b): gain $G$ as a function of the squeezing parameter $s$ for our noisy detector (solid red line) and in the ideal case (dashed blue line). 
Here $\mean{N}=(\Delta N)^2 = 36$.
Panel (c): gain as a function of $\mean{N}$, when optimized over $s$ and for fixed $\Delta N =6$.
In all panels the gain is shown in linear scale.}  
\label{FigSupp3}
\end{figure*}

Given the state $\hat{\rho}$, Eq.~(\ref{rhostate}), we study the sensitivity to phase shift estimation when applying the transformation $e^{-i \theta \hat{J}_y}$. 
The sensitivity is calculated via error propagation when measuring the number of particles in one atomic level only, namely
\be
(\Delta \theta)^2 = \frac{(\Delta N)^2}{(d\langle N \rangle/d\theta)^2},
\ee
eventually taking into account the finite detection resolution.
When optimized over $\theta$, the sensitivity depends strongly on the squeezing parameter as well as the finite width of $\rho_N$. 
In Fig.~\ref{FigSupp3}(a) we plot the sensitivity gain over the shot noise, namely $G=1/ [\mean{N} (\Delta \theta)^2]$ as a function of $\Delta N$ and $s$.
Here we consider $\mean{N}\approx 36$, as in the experiment, and use the reconstructed experimental detection. 
The white region corresponds to $G\leq 1$. 
We see that the gain increases when decreasing $\Delta N$ and $s$, and can reach $G \approx 6$ for our relatively small number of particles and finite detection resolution. 

In Fig.~\ref{FigSupp3}(b), we plot the gain as a function of $s$, for $\mean{N}\approx 36$ and fixed $(\Delta N)^2 = 36$, as in the experiment. 
The blue line is the gain obtained with ideal detection, while the red line is obtained for our noisy detector (namely obtained by using our reconstructed detector matrix $V$).
For $s=1$ (no squeezing), we find $G \approx 0.75$, consistent with the analysis of the Fisher information shown in Fig.~\ref{FigSupp2}.
In Fig. \ref{FigSupp3}(c) we plot the gain as a function of $\mean{N}$, optimized over $s$, and for $\Delta N = \sqrt{\mean N}$.  
The blue line (obtained for ideal detection) follows $G \sim \bar{N}^{0.25}$. 
The red line (obtained for our noisy detection) tends to approach the blue line for $\mean{N}\gg1$. 

\end{document}